# Heavy-ion fusion cross sections of weakly bound $^9$Be on $^{27}$Al, $^{64}$Zn and tightly bound $^{16}$O on $^{64}$Zn target using Coulomb and proximity potential

K. P. Santhosh and V. Bobby Jose

School of Pure and Applied Physics, Kannur University, Swami Anandatheertha Campus, Payyanur 670327, INDIA.

**Abstract**

The total fusion cross sections for the fusion of weakly bound $^9$Be on $^{27}$Al and $^{64}$Zn targets at near and above the barrier have been calculated using one dimensional barrier penetration model, taking scattering potential as the sum of Coulomb and proximity potential and the calculated values are compared with experimental data. For the purpose of comparison of the fusion of weakly bound projectiles and strongly bound projectiles, the total fusion cross sections for the reaction of tightly bound nucleus $^{16}$O on $^{64}$Zn have also been computed using a similar procedure. The calculated values of total fusion cross sections in all cases are compared with coupled channel calculations using the code CCFULL. The computed cross sections using Coulomb and proximity potential explain the fusion reactions well in both cases of weakly bound and strongly bound projectiles. Reduced reaction cross sections for the systems $^9$Be+$^{27}$Al, $^9$Be+$^{64}$Zn and $^{16}$O+$^{64}$Zn have also been described.

## 1. Introduction

As the synthesis of super heavy elements (SHEs) is a hot topic and very interesting problem nowadays, many investigations both experimental and theoretical on heavy-ion fusion reactions at, above and below the Coulomb barrier [1-5] have been an area of extensive studies in Nuclear Physics. The considerably high beam intensity of stable weakly bound nuclei compared to radioactive beams make the study of the fusion of weakly bound nuclei and the influence of breakup process in the fusion reactions [6-11] an important subject of detailed and in-depth investigation in heavy-ion fusion reactions. In the fusion processes and more specifically, in the fusion of weakly bound nuclei, several reaction mechanisms followed by the breakup of the projectile have to be considered both experimentally and theoretically: incomplete fusion reaction (ICF), when part of the projectile is captured by the target; and complete fusion (CF) with the capture of all of the projectile constituents by the target. Total fusion (TF) is understood as the sum of these two processes (CF+ICF).

In the analysis of heavy-ion fusion reactions, an internuclear interaction consisting of repulsive Coulomb and centrifugal potentials and attractive nuclear potential plays a major role, where the potential is a function of the distance between centres-of mass of the colliding nuclei. At a distance referred to as Coulomb barrier the total potential attains a maximum value, where the repulsive and attractive forces balance each other and the energy of relative motion must overcome this barrier in order for the nuclei to be captured and fused.

In the present work, the fusion excitation functions for the fusion of weakly bound $^9$Be on $^{27}$Al and $^{64}$Zn targets have been calculated using one dimensional barrier penetration model [1], taking scattering potential as the sum of Coulomb and proximity potential [12] and the calculated values are compared with experimental data [11,13,14]. For the purpose of comparison of the fusion of weakly bound projectiles and strongly bound projectiles, the total fusion cross sections for the reaction of tightly bound nucleus $^{16}$O on $^{64}$Zn have also been computed using a similar procedure and the results are compared with experimental data [15]. The calculated values of total fusion cross sections in all cases are compared with coupled channel calculations using the code CCFULL [16]. Reduced reaction cross sections for the systems $^9$Be+$^{27}$Al, $^9$Be+$^{64}$Zn and $^{16}$O+$^{64}$Zn have been described, by using the usual reduction procedure of dividing the cross section by $\pi R_0^2$, where $R_0$ is the barrier radius and the division of energy by Coulomb barrier.

## 2. Theory
### 2.1. The potential

Describing a unique nuclear potential for exploring different nuclear reaction mechanisms, which are exclusively governed by the nucleus-nucleus potential is an extensive challenge for the last several years in nuclear physics. Explaining the nuclear potential as the product of a geometrical factor, which is proportional to the reduced radii of colliding nuclei and a universal function is commonly accepted, as it is incorporating the role of different colliding nuclei in the geometrical factor. In this effort, a simple formula for the nucleus-nucleus interaction energy as a function of separation between the surfaces of the approaching nuclei has been given by the proximity potential of Blocki et al. [17]. The formula is free of adjustable parameters and makes use of the measured values of the nuclear surface tension and surface diffuseness.

The interaction barrier for two colliding nuclei is given as:

$$V = \frac{Z_1 Z_2 e^2}{r} + V_P(z) + \frac{\hbar^2 \ell(\ell+1)}{2\mu r^2} \tag{1}$$

where $Z_1$ and $Z_2$ are the atomic numbers of projectile and target, r is the distance between the centers of the projectile and target, z is the distance between the near surfaces of the projectile and target, $\ell$ is the angular momentum, $\mu$ is the reduced mass of the target and projectile and $V_P(z)$ is the proximity potential given as:

$$V_P(z) = 4\pi\gamma b \frac{C_1 C_2}{C_1 + C_2} \phi(\frac{z}{b}) \tag{2}$$

with the nuclear surface tension coefficient,

$$\gamma = 0.9517[1 - 1.7826(N-Z)^2/A^2] \tag{3}$$

$\phi$, the universal proximity potential is given as:

$$\phi(\xi) = -4.41\exp(-\xi/0.7176), \quad \text{for } \xi \geq 1.9475 \tag{4}$$

$$\phi(\xi) = -1.7817 + 0.9270\xi + 0.01696\xi^2 - 0.05148\xi^3, \quad \text{for } 0 \leq \xi \leq 1.9475 \tag{5}$$

$$\phi(\xi) = -1.7817 + 0.9270\xi + 0.0143\xi^2 - 0.09\xi^3, \quad \text{for } \xi \leq 0 \tag{6}$$

with $\xi = z/b$, where the width (diffuseness) of nuclear surface $b \approx 1$ and Süssmann Central radii $C_i$ related to sharp radii $R_i$ as $C_i = R_i - \frac{b^2}{R_i}$. For $R_i$, we use the semi empirical formula in terms of mass number $A_i$ as:

$$R_i = 1.28 A_i^{1/3} - 0.76 + 0.8 A_i^{-1/3} \tag{7}$$

The choice of the potential and its form to be adopted is one of the most challenging aspects, when one wants to compare the experimental fusion data with theory. Among such potentials, proximity potential is well known for its simplicity and numerous applications in different fields. It is based on the proximity force theorem according to which the nuclear part of the interaction potential is a product of the geometrical factor depending on the mean curvature of the interaction surface and the universal function (depending on the separation distance) and is independent of the masses of the colliding nuclei.

## 2.2. The fusion cross section

For the last four decades, the barrier penetration model developed by C. Y. Wong [1] has been widely used to describe the fusion reactions at energies not too much above the barrier and at higher energies, which obviously explains the experimental result properly. Following Thomas [18], Huizenga and Igo [19] and Rasmussen and Sugawara [20], Wong approximated the various barriers for different partial waves by inverted harmonic oscillator potentials of height $E_\ell$ and frequency $\omega_\ell$. For energy $E$, using the probability for the absorption of $\ell^{th}$ partial wave given by Hill-Wheeler formula [21], Wong arrived at the total cross section for the fusion of two nuclei by quantum mechanical penetration of simple one-dimensional potential barrier as:

$$\sigma = \frac{\pi}{k^2} \sum_\ell \frac{2\ell+1}{1+\exp[2\pi(E_\ell - E)/\hbar\omega_\ell]} \tag{8}$$

where $k = \sqrt{\frac{2\mu E}{\hbar^2}}$. Here $\hbar\omega_\ell$ is the curvature of the inverted parabola. Using some parameterizations in the region $\ell = 0$ and replacing the sum in Eq. (8) by an integral Wong gave the reaction cross section as:

$$\sigma = \frac{R_0^2 \hbar\omega_0}{2E} \ln\left\{1+\exp\left[\frac{2\pi(E-E_0)}{\hbar\omega_0}\right]\right\} \tag{9}$$

For relatively large values of $E$, the above result reduces to the well-known formula:

$$\sigma = \pi R_0^2 \left[1 - \frac{E_0}{E}\right] \tag{10}$$

## 2.3. The reduced reaction cross section

A common procedure of eliminating the geometrical factors concerning different systems by 'reducing' the cross section and the centre-of-mass energy has extensively been used in recent years [22, 23,24] for the comparison of the excitation functions of different reaction mechanisms induced by different projectiles on the same target nucleus. The normal procedure consists of the division of the cross section by $\pi R_0^2$, where $R_0$ is the barrier radius and the division of energy by Coulomb barrier $E_0$.

## 3. Results and discussions

For $\ell = 0$, the interaction barriers for the fusion of the systems $^9$Be+$^{27}$Al, $^9$Be+$^{64}$Zn and $^{16}$O+$^{64}$Zn with Coulomb and proximity potential have been plotted in Fig.1, against the distance between the centers of the projectile and the target. The values of the barrier height $E_0$ and the barrier radius $R_0$ for the reactions are noted. It is to be noted that the system $^{16}$O+$^{64}$Zn has maximum barrier height $E_0$ and the barrier radius $R_0$. In the case of reactions of weakly bound projectile $^9$Be on $^{27}$Al and $^{64}$Zn, as the mass of the target increases, the barrier height $E_0$ increases and the barrier radius $R_0$ shifts towards larger value. Above the Coulomb barrier, the total fusion cross-sections for all the reactions have been calculated by using the values of barrier height $E_B$ and barrier radius $R_B$ taken from the respective figures and using Eq. (8).

In the reactions of $^9$Be+$^{17}$Al, $^9$Be+$^{64}$Zn and $^{16}$O+$^{64}$Zn, above the barrier, the total fusion cross-sections computed using Wong's formula with Coulomb and proximity potential given by Eq. (8) and the corresponding experimental values are shown in the Tables I, II and III and in Figs. 2, 3 and 4.

In Figs. 2, 3 and 4, the dotted lines represent calculations using no channel couplings and the solid lines represent the vibrational couplings in the target, while using the code CCFULL, whereas pair transfer channel is not included. A radius parameter of $r_c$=1.2fm, with deformation parameters, $\beta_\lambda^N = \beta_\lambda^C$ is used in all calculations. The depth parameter $V_0$ and the surface diffuseness parameter $a_0$ for the nuclear potential (of the Wood-Saxon form) in the coupled channel calculations using CCFULL have been calculated [25], with $r_0$ fixed at 1.20fm.

In Fig.2, in the case of reaction of weakly bound projectile $^9$Be on $^{27}$Al target, above the barrier the total fusion cross-sections (down triangles) computed using Eq.(8) fit reasonably well with the experimental data (circles), while using the scattering potential as Coulomb and proximity potential. The depth parameter $V_0$ and the surface diffuseness parameter $a_0$ for the nuclear potential in CCFULL calculations have been taken as 40.40MeV and 0.60fm respectively. In the CCFULL calculations (Coupled) of $^9$Be+$^{27}$Al reaction, the single phonon excitation with $\beta_2^N = 0.448$ and excitation energy 1.014MeV of $^{27}$Al is used, where the weakly bound projectile $^9$Be is kept inert. In the case of CCFULL calculations with no couplings (Uncoupled) almost no difference between the two calculations can be observed in comparison with coupled calculations, as expected at these energies above the Coulomb barrier.

In the case of reaction of weakly bound $^9$Be and strongly bound $^{16}$O projectiles on $^{64}$Zn target, above the barrier the total fusion cross-sections (down triangles) computed using Eq.(8)

show good fit with the experimental data (circles), irrespective of the fact that $^9$Be projectile is weakly bound and $^{16}$O is strongly bound, as shown in Figs.3 and 4. The dotted lines and the solid lines in the Figs. 3 and 4 show the CCFULL calculations without and with coupling respectively. In all CCFULL calculations, the single phonon excitations in $^{64}$Zn with $\beta_2^N = 0.236$ and the first two exited states of the target with excitation energies 1.799MeV and 0.991MeV have been used, where the projectiles are kept inert. The ($V_0$, $a_0$) parameters for the reactions $^9$Be+$^{64}$Zn and $^{16}$O+$^{64}$Zn in the CCFULL calculations have been taken as (46.23MeV, 0.62fm) and (54.95MeV, 0.64fm) respectively. In the case of with and without couplings, almost no difference between the two calculations can be observed just as it is expected above the Coulomb barrier.

The comparatively good agreement between all calculations with the experimental data, at least within the experimental uncertainties, it can be concluded that the total fusion cross section calculations for the systems are not affected by the breakup process at energies above the barrier while using Coulomb and proximity potential.

If the computed fusion cross sections for the system $^{16}$O+$^{64}$Zn with strongly bound projectile is directly compared with the system $^9$Be+$^{64}$Zn with loosely bound projectile, contribution of the geometrical factors such as size of the projectile influences the result. For eliminating the effect, we have renormalized the cross sections with respect to geometrical value and incident energy with respect to barrier height, for the systems $^9$Be+$^{64}$Zn and $^{16}$O+$^{64}$Zn. Fig. 5 represents the corresponding reduced reaction cross sections, by the usual reduction procedure of dividing the cross section by $\pi R_0^2$ and the energy by $E_0$ so as to compare the two systems on the same figure. As the cross sections for the two systems are rather similar, as shown in Fig.5, it can be concluded that the static effects arising from the weakly bound nucleons do not affect the fusion cross sections above the barrier and moreover, the dynamic channel couplings also are not relevant above the barrier.

## 4. Conclusions

Interaction barriers for the fusion of the systems $^9$Be+$^{27}$Al, $^9$Be+$^{64}$Zn and $^{16}$O+$^{64}$Zn $^9$Be+$^{64}$Zn have been plotted against the distance between the centers of the projectile and target. It is noted that the barrier height $E_0$ and the barrier radius $R_0$ are greater for the system $^{16}$O+$^{64}$Zn. The total reaction cross sections of the systems $^9$Be+$^{27}$Al, $^9$Be+$^{64}$Zn and $^{16}$O+$^{64}$Zn have also been computed and the results are compared with experimental data and coupled channel calculations using code CCFULL. While using the scattering potential as the sum of Coulomb and proximity potentials, above the barrier, the simple one dimension barrier penetration model developed by C. Y. Wong explains the fusion reactions of heavy ions reasonably well, irrespective of the nature of the projectiles; either weakly bound or strongly bound. The reduced cross sections compare different fusion reaction mechanisms induced by different projectiles and targets in the same figure. As the reduced reaction cross sections for the systems $^9$Be+$^{64}$Zn and $^{16}$O+$^{64}$Zn are rather similar, it can be concluded that the static effects arising from the weakly bound nucleons do not affect the fusion cross sections above the barrier and moreover, the dynamic channel couplings also are not relevant above the barrier.

**References**

 [1] C. Y. Wong, Phys. Rev. Lett. **31**, 766 (1973)
 [2] N. Rowley, G. R. Satchler and P. H. Stelson, Phys. Lett. B **254**, 25 (1991)
 [3] R. K. Puri and R. K. Gupta, Phys. Rev. C **45**, 1837 (1992)
 [4] J. R. Leigh, M. Dasgupta, D. J. Hinde, J. C. Mein, C. R. Morton, R. C. Lemmon, J. P. Lestone, J. O. Newton, H. Timmers, J. X. Wei and N. Rowley, Phys. Rev. C **52**, 3151 (1995)
 [5] K. P. Santhosh, V. Bobby Jose, Antony Joseph and K. M. Varier, Nucl. Phys. A **817**, 35 (2009)
 [6] N. Takigawa and H. Sagawa, Phys. Lett. B **265**, 23 (1991)
 [**7**] M. S. Hussein, M. P. Pato, L. F. Canto, and R. Donangelo, Phys. Rev. C **46**, 377 (1992)
 [8] N. Takigawa, M. Kuratani, and H. Sagawa, Phys. Rev. C **47**, 2470(R) (1993)
 [9] P. R. S. Gomes, J. Lubian, and R. M. Anjos, Nucl. Phys. A **734**, 233 (2004)
[10] M. S. Hussein, L. F. Canto, and R. Donangelo, Nucl. Phys. A **722**, 321 (2003)
[11] P. R. S. Gomes, M. D. Rodriguez, G. V. Marti, I. Padron, L. C. Chamon, J. O. Fernandez Niello, O. A. Capurro, A. J. Pacheco, J. E. Testoni, A. Arazi, M. Ramirez, R. M. Anjos, J. Lubian, R. Veiga, R. Liguori Neto, E. Crema, N. Added, C. Tenreiro, and M. S. Hussein Phys. Rev. C **71**, 034608 (2005)
[12] K. P. Santhosh and A. Joseph, Pramana J. Phys. **58**, 611 (2002)
[13] G. V. Martí, P. R. S. Gomes, M. D. Rodríguez, J. O. Fernández Niello, O. A. Capurro, A. J. Pacheco, J. E. Testoni, M. Ramírez, A. Arazi, I. Padron, R. M. Anjos, J. Lubian and E. Crema, Phys. Rev. C **71**, 027602 (2005)

[14] S. B. Moraes, P. R. S. Gomes, J. Lubian, J. J. S. Alves, R. M. Anjos, M. M. Sant'Anna, I. Padron, C. Muri, R. Liguori Neto, and N. Added, Phys. Rev. C **61**, 064608 (2000)

[15] C. Tenreiro et al., in Proceedings of the Workshop on Heavy Ion Fusion: Exploring the Variety of Nuclear Properties, Padova, Italy, 1994, edited by A. Stefanini (World Scientific, Singapore1994), p. 98

[16] K. Hagino, N. Rowley and A. T. Kruppa, Comput. Phys. Commun. **123**, 143 (1999)

[17] J. Blocki, J. Randrup, W. J. Swiatecki and C. F. Tsang, Ann. Phys. (N.Y) **105**, 427 (1977)

[18] T. D. Thomas, Phys. Rev. **116**, 703 (1959)

[19] J. Huizenga and G. Igo, Nucl. Phys. **29**, 462 (1961)

[20] J. Rasmussen and K. Sugawara-Tanabe, Nucl. Phys. A **171**, 49620 (1971)

[21] D. L. Hill and J. A. Wheeler, Phys. Rev. **89**, 1102 (1953)

[22] M. Beckerman, M. Salomaa, A. Sperduto, J. D. Molitoris and A. DiRienzo, Phys. Rev. C **25**, 837 (1982)

[23] D. E. DiGregorio, J. O. Fernandez Niello, A. J. Pacheco, D. Abriola, S. Gil, A. O. Macchiavelli, J. E. Testoni, P. R. Pascholati, V. R. Vanin, R. Liguori Neto, N. Carlin Filho, M. M. Coimbra, P. R. S. Gomes and R. G. Stokstad, Phys. Lett. B **176**, 322 (1986)

[24] Canto L F, Gomes P R S, Lubian J, Chamon L C and Crema E 2009 Nucl. Phys. A **821**, 51

[25] Winther A 1995 Nucl.Phys.A **594** 203

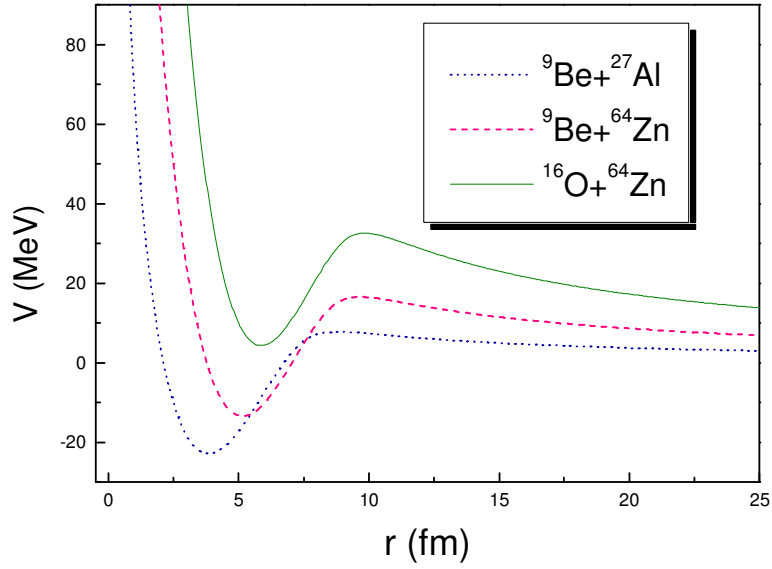

**Figure 1.** (Color Online) Scattering potential for the reactions of $^9$Be + $^{27}$Al, $^9$Be + $^{64}$Zn and $^{16}$O + $^{64}$Zn consisting of repulsive Coulomb and centrifugal potentials and attractive nuclear proximity potential, for $\ell = 0$.

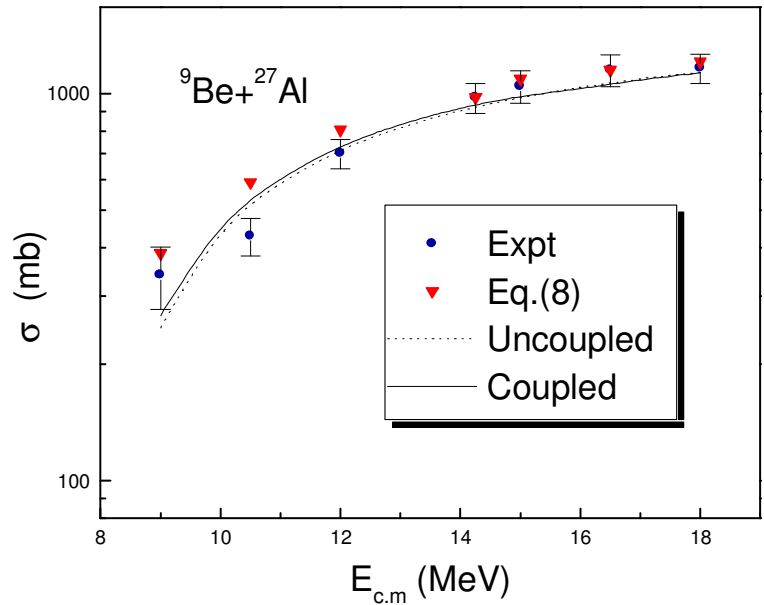

**Figure 2.** (Color Online) Computed fusion cross sections for $^9$Be + $^{27}$Al reaction using Wong's formula and their comparison with experimental data and CCFULL.

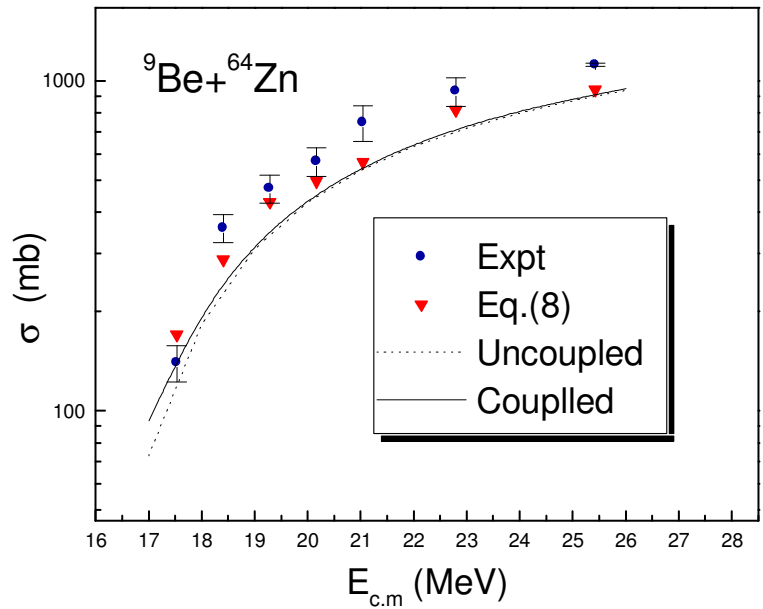

**Figure 3.** (Color Online) Computed fusion cross sections for $^9$Be + $^{64}$Zn reaction using Wong's formula and their comparison with experimental data and CCFULL.

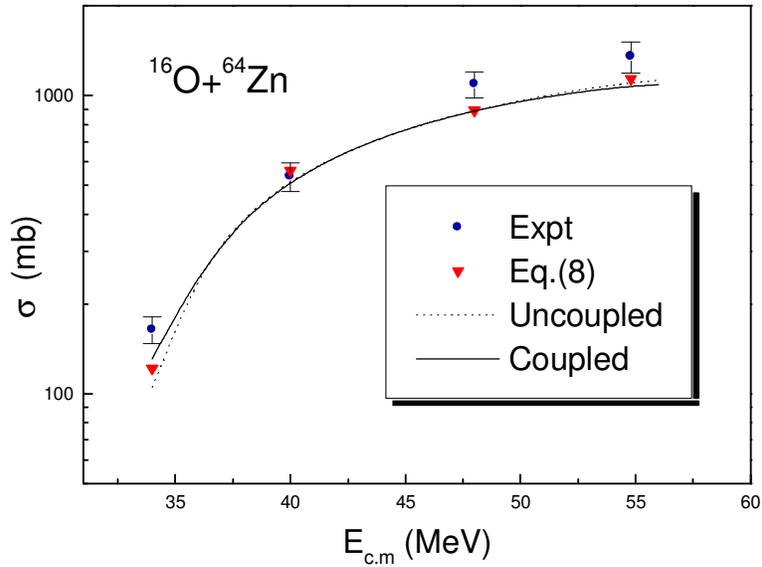

**Figure 4.** (Color Online) Computed fusion cross sections for $^{16}$O + $^{64}$Zn reaction using Wong's formula and their comparison with experimental data and CCFULL.

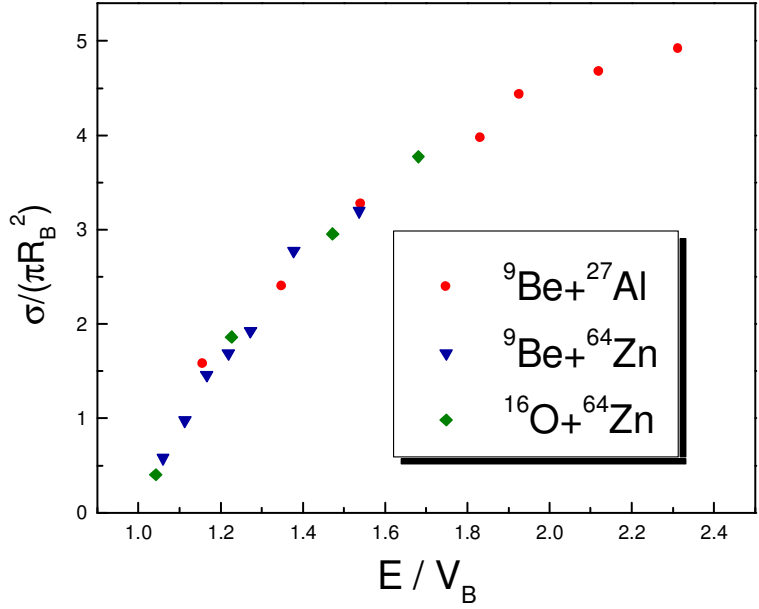

**Figure 5.** (Color Online) Reduced reaction cross sections for the systems $^9$Be + $^{27}$Al, $^9$Be + $^{64}$Zn and $^{16}$O + $^{64}$Zn.

**Table 1.** Computed fusion cross sections for the system $^9$Be + $^{27}$Al their comparison with experimental data.

| $E_{c.m}$ (MeV) | Fusion cross sections (mb) | |
|---|---|---|
| | Expt. Ref. [13] | Theory Eq.(8) |
| 9.00 | 339±62 | 386.35 |
| 10.50 | 428±48 | 588.73 |
| 12.00 | 700±61 | 804.91 |
| 14.25 | 975±88 | 976.06 |
| 15.00 | 1044±100 | 1088.24 |
| 16.50 | 1149±108 | 1147.36 |
| 18.00 | 1163±100 | 1207.37 |

**Table 2.** Computed fusion cross sections for the system $^9$Be + $^{64}$Zn their comparison with experimental data.

| $E_{c.m}$ (MeV) | Fusion cross sections (mb) | |
|---|---|---|
| | Expt. Ref. [11,14] | Theory Eq.(8) |
| 17.53 | 140±18 | 169.68 |
| 18.41 | 358±35 | 287.25 |
| 19.28 | 472±46 | 428.42 |
| 20.16 | 570±57 | 495.84 |
| 21.04 | 747±92 | 565.49 |
| 22.79 | 930±92 | 815.63 |
| 25.42 | 1120±112 | 939.26 |

**Table 3.** Computed fusion cross sections for the system $^{16}$O + $^{64}$Zn and their comparison with experimental data.

| $E_{c.m}$ (MeV) | Fusion cross sections (mb) | |
|---|---|---|
| | Expt. Ref. [15] | Theory Eq.(8) |
| 34.00 | 164±17 | 121.34 |
| 40.00 | 536±60 | 561.56 |
| 48.00 | 1095±110 | 892.44 |
| 54.80 | 1354±162 | 1138.62 |